# Content Base Image Retrieval Using Phong Shading

UDAY PRATAP SINGH,
Assistant Professor of CSE Department.
LNCT, Bhopal (M.P) INDIA
usinghiitg@gmail.com

SANJEEV JAIN,
Professor & Head of CSE Department
LNCT, Bhopal (M.P) INDIA
dr__sanjeevjain@yahoo.com

GULFISHAN FIRDOSE AHMED,
M.Tech. Student of CSE Department.
LNCT, Bhopal (M.P) INDIA
gul_firdose@rediffmail.com

*Abstract:-*The digital image data is rapidly expanding in quantity and heterogeneity. The traditional information retrieval techniques does not meet the user's demand, so there is need to develop an efficient system for content based image retrieval. Content based image retrieval means retrieval of images from database on the basis of visual features of image like as color, texture etc. In our proposed method feature are extracted after applying Phong shading on input image. Phong shading, flattering out the dull surfaces of the image The features are extracted using color, texture & edge density methods. Feature extracted values are used to find the similarity between input query image and the data base image. It can be measure by the Euclidean distance formula. The experimental result shows that the proposed approach has a better retrieval results with phong shading.

## I. INTRODUCTION

In recent years, Content Based Image Retrieval (CBIR) has played an important role in many fields, such as medicine, geography, security. General approaches in CBIR are based on visual attributes of images such as color, texture, shape etc. Most of CBIR systems are designed to find the top N images that are most similar to the input query image [2]. The most common categories of descriptors are based on color, texture and shape. An efficient image retrieval system must be based on efficient image feature descriptors. Image retrieval methods may also depend on the properties of the images being analyzed [3]. The purpose of this paper is to develop a CBIR system using the concept of phong shading. The proposed solution is to first apply the phong shading on query image to enhance the visual quality and then extract the visual features of query image and compare them to the database features used are color texture & edge density. The mean, median and standard deviation of red, green and blue channels of color histogram. The texture features such as contrast, energy & entropy are retrieved from phong shaded gray scale image. The edge feature includes vertical and horizontal edges Euclidean distance formulas have been used for similarity measurement [1].

## II. PROPOSED FRAMEWORK

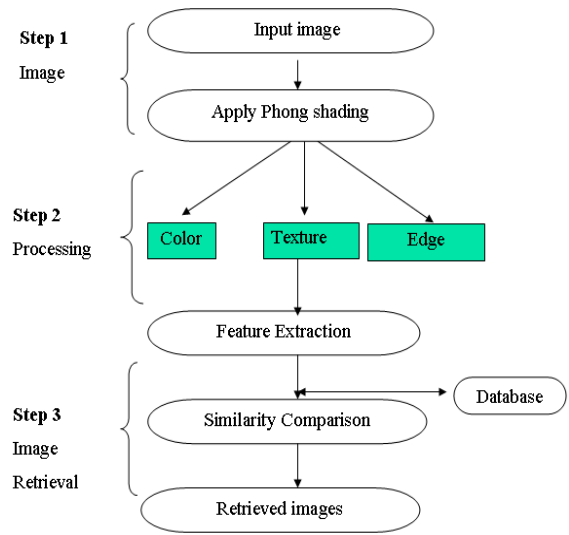

Figure 4.1 Proposed Framework

Proposed frame work is as follows:-

In proposed framework first apply the phong shading on each image of database then extract its color texture & edge features and then construct single feature vector for each image. Feature database has feature vector of each image and same procedure is applied on query image to extract its features then compare its feature vector with feature database to retrieve the most similar images.

## II. PHONG SHADING

The interpolation method may also be called Phong interpolation, which is usually referred to by "per-pixel lighting". Typically it is called "shading".
An illumination model define as model for determining a radiant intensity at a given point in space, for single light source it can be formulated as follows:

$$I = K_a * I_a + K_d * I_l * (\vec{N} \bullet \vec{L}) + K_s * I_l * (\vec{N} \bullet \vec{H})^{N_S} \quad (1)$$

This formula states that the intensity at a certain point depends on the reflectance of the object ($K_a$, $K_d$ and $K_s$ are ambient, diffuse, and specular reflectance, respectively), on the intensities of the light source $I_l$, and on some vectors. $\vec{N}$ denotes the unit normal vector to the surface. $\vec{L}$ denotes the unit light source vector, and $\vec{H}$ is the unit halfway vector indicating the direction of the maximum highlight. Another object dependent feature is expressed by the glossiness $N_s$ which ranges from one to infinity. Phong shading is a well known method for producing realistic shading but it has not been used by real-time systems The phong shading is basically used to enhance the visual quality of an image to extract features more accurately. The Phong model describes the interaction of light with a surface, in terms of the properties of the surface and the nature of the incident light. The Phong model reflected light in terms of a diffuse and specular component together with an ambient term. The intensity of a point on a surface is taken to be the linear combination of these three components. Phong shading calculates the average unit normal vector at each of the polygon vertices and then interpolates the vertex normal over the surface of the polygon after calculating the surface normal it applies the illumination model to get the color intensity for each pixel of the polygon.

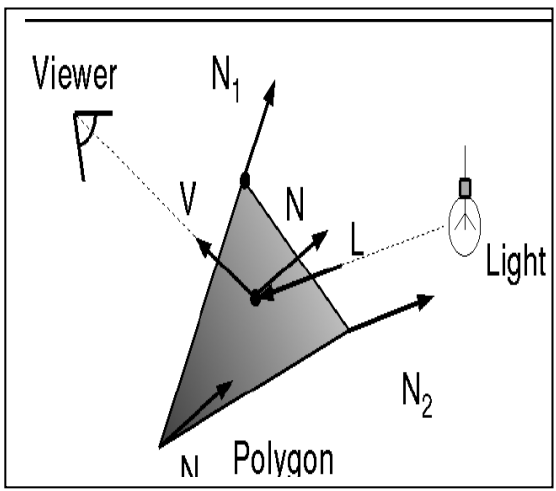

Figure2- Phong shading (interpolation of vectors)

The intensity of light on the surface is the combination of diffuse reflection, ambient light & specular reflection the intensity of diffused light is given in fig.2

For phong shading $\vec{N} \bullet \vec{H}$ of the specular term can be formulated as

$$\vec{N} \bullet \vec{H} = \frac{\vec{N}_{nn} \bullet \vec{H}_{nn}}{\vec{N}_{nn} \left| \vec{H}_{nn} \right|}$$

$$= \frac{\left(\vec{A}x + \vec{B}y + \vec{C}\right) \bullet \left(\vec{D}x + \vec{E}y + \vec{F}\right)}{\left(\vec{A}x + \vec{B}y + \vec{C}\right) \left| \left(\vec{D}x + \vec{E}y + \vec{F}\right) \right|} \quad (2)$$

where $\vec{N}_{nn} = \vec{A}x + \vec{B}y + \vec{C}$ represents the approximate nonunit length normal vector that is linearly interpolated from the true surface normal vectors specified at the vertices. $\vec{A}, \vec{B}, \vec{C}$ are chosen to interpolate the surface normal vector across the polygon. $\vec{H}_{nn} = \vec{D}x + \vec{E}y + \vec{F}$ represents the nonunit length halfway vector between the light source vector and viewer vector.

### III. COLOR FEATURES

One of the most important features that make possible the recognition of images by humans is a color. Color is a property that depends on the reflection of light to the eye and the processing of that information in the brain. The main method of representing color information of images in CBIR systems is through color histograms. For extracting the color features, the color histogram is used as it is independent of image size & orientation. Here the RGB color histogram is used. for extracting these feature first we find the histogram for red, green & blue channel..[1, 3, 4, 5, 10]

### IV. TEXTURE FEATURES

Texture is that innate property of all surfaces that describes visual patterns, each having properties of homogeneity. It contains important information about the structural arrangement of the surface, such as; clouds, leaves, bricks, fabric, etc. The texture features describes the relationship of the surface to the surrounding environment. A texture is characterized by a set of values called energy, entropy, contrast, and homogeneity.In short, it is a feature that describes the distinctive physical composition of a surface.The following formulas are used to calculate the features and are shown in equations 3 to 6 [1, 9]:-

A. *Entropy:* -.
Entropy is defined as

$$Entropy = \sum_i \sum_j P(i,j) \log P(i,j) \quad (3)$$

Where P contains the histogram counts.

*B. Contrast:-*

It is the measure of the intensity contrast between a piel and its neighbor over the whole image.

$$Contrast = \sum_i \sum_j (i-j)^2 P(i,j) \quad (4)$$

*C. Energy: -*

It is the sum of squared elements.

$$Energy = \sum_i \sum_j P^2(i,j) \quad (5)$$

*D. Homogeneity: -*

A value that measures the closeness of the distribution of elements.

$$Homogeneity = \sum_i \sum_j \frac{P(i,j)}{1+|i-j|} \quad (6)$$

## V. EDGE DETECTION

Edge detection refers to the process of identifying and locating sharp discontinuities in an image. The discontinuities are abrupt changes in pixel intensity which characterize boundaries of objects in a scene. Edge detection is a terminology in image processing and computer vision, particularly in the areas of feature detection and feature extraction, to refer to algorithms which aim at identifying points in a digital image at which the image brightness changes sharply or more formally has discontinuities. Edge detection would then be done on the intensity channel of a color image in HSI space. Another definition claims an edge exists if it is present in the red green and the blue channel. Performing it on each of the color component can do edge detection.Edge detection refers to the process of identifying and locating sharp discontinuities in an image. The discontinuities are abrupt changes in pixel intensity which characterize boundaries of objects in a scene. [1, 10]

*A. Sobel Edge detection:-*

The Sobel operator is used in image processing, particularly within edge detection algorithms. The Sobel operator is based on convolving the image with a small, separable, and integer valued filter in horizontal and vertical direction and is therefore relatively inexpensive in terms of computations. The operator consists of a pair of 3×3 convolution kernels mask as shown below. One kernel is simply the other rotated by 90°.

| -1 | 0 | +1 |
|----|---|----|
| -2 | 0 | +2 |
| -1 | 0 | +1 |

Gx

| +1 | +2 | +1 |
|----|----|----|
| 0  | 0  | 0  |
| -1 | -2 | -1 |

Gy

## VI. SIMILARITY COMPARISON

CBIR employs low level image features such as color, shape or texture to achieve objective and automatic indexing, in contrast to subjective and manual indexing in traditional image indexing. For contend based image retrieval, the image feature extracted is usually an N-dimensional feature vector which can be regarded as a point in a N-dimensional space. A similarity measurement is normally defined as a metric distance.For similarity comparison we have used the Euclidean distance formula. Two pixels has coordinates $(X_1, Y_1)$ and $(X_2, Y_2)$ then the Euclidean distance is given by:

$$D_{Euclid} = \sqrt{(X_2 - X_1)^2 + (Y_2 - Y_1)^2} \quad (7)$$

## VII. EXPERIMENTAL RESULT AND ANALYSIS

For retrieval efficiency and effect of phong shading on an image. From the corel database 70 images with five different subjects each contain 14 images are taken. Calculate a recall & precision value in both cases output with phong shading and without phong shading on an image. Recall and precision depend on the outcome of a query and all its relevant and non-relevant images.
Standard formulas have been used to compute the precision and recall for query images is shown below.

Precision is defined as the number of relevant documents retrieved by a search divided by the total number of documents retrieved by that search.

$$precision = \frac{No.of\ relevent\ images\ retrieved}{Total\ no.of\ images\ retrieved}$$

Recall is defined as the number of relevant documents retrieved by a search divided by the total number of existing relevant documents (which should have been retrieved).

$$Recall = \frac{No\ of\ relevant\ images\ retrieved}{Total\ no.\ of\ relevant\ images\ in\ the\ database}$$

For calculating recall & precision value we have consider database of 70 images randomly selected from Caltech_256 image database. In our database we have consider 5 different categories and each categories contain 14 image each.

Fig3 (without phong shaded) and Fig4(with phong shaded) is shown below. Feature value becomes are same but in RGB colour histogram, value and range both are different in both cases.

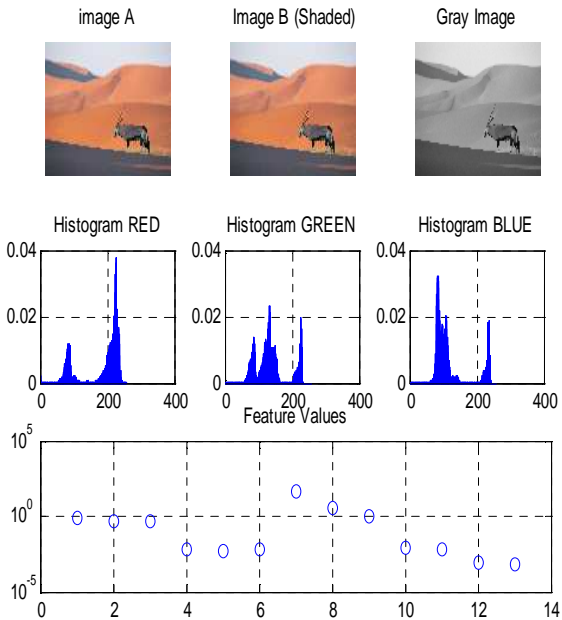

Figure3: Image without phong shading

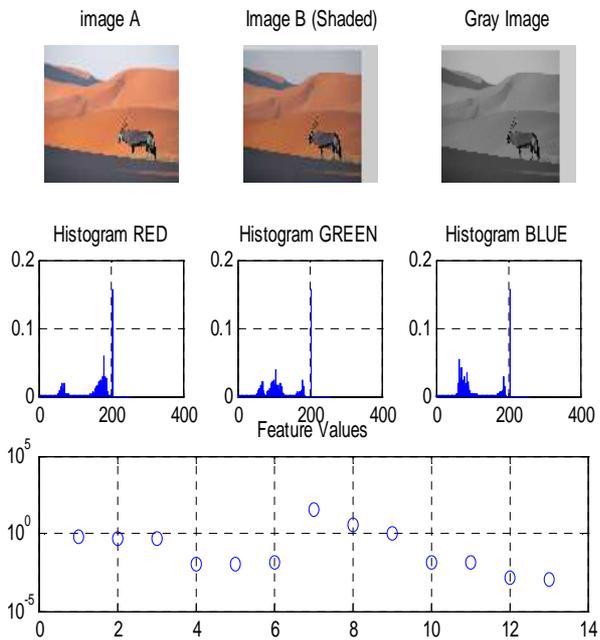

Figure4: Image with phong shading

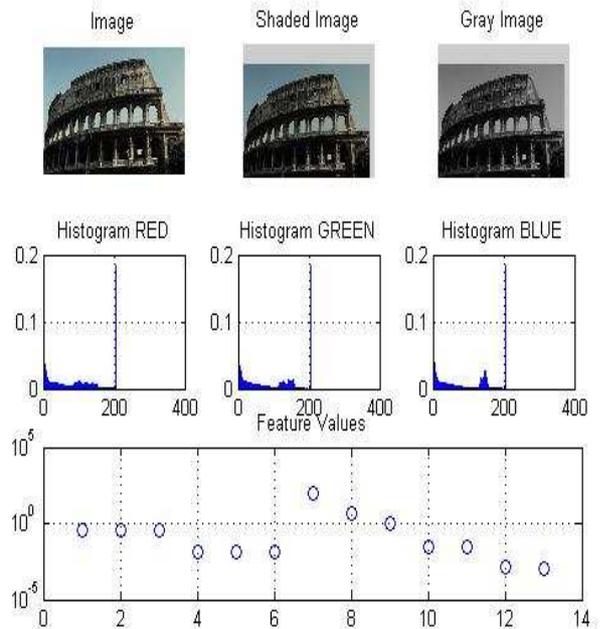

Figure5: Query image it's shaded & gray image

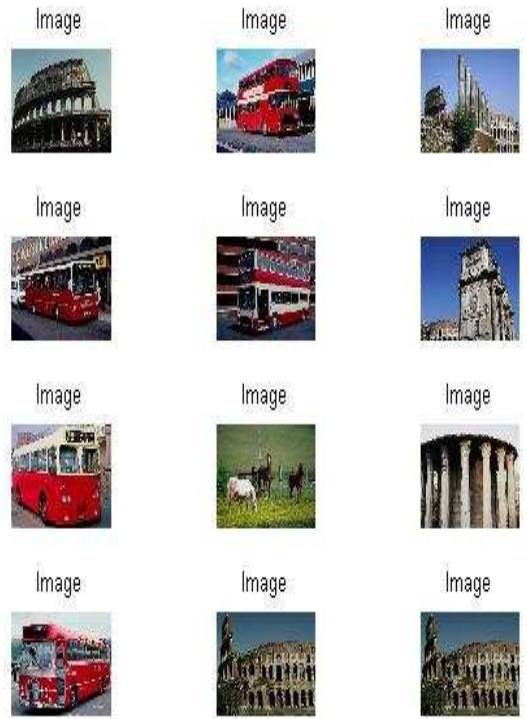

Figure6: Output of Query image with phong shading

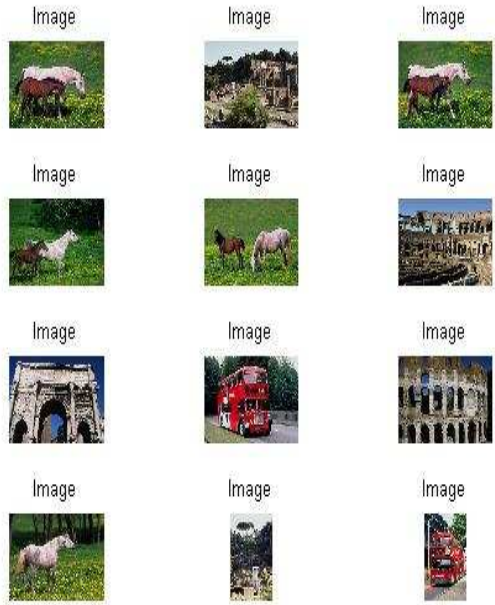

Figure7: Output of Query image without phong shading

For result analysis we have consider the following five different categories of images:-

Category 1= Historical Buildings
Category 2= Buses
Category 3= Dianasour
Category 4= Flowers
Category 5 = Horse

Calculated recall and precision value of selected input image for with and without phong shaded are shown in Table1, Table2, Fig.8 and Fig.9.

TABLE I.: Precision and Recall value with Phong Shading

| Category | No. of relevant image retrieved | Precision (%) | Recall (%) |
|---|---|---|---|
| 1 | 6 | 50% | 43% |
| 2 | 6 | 50% | 43% |
| 3 | 10 | 83% | 71% |
| 4 | 9 | 75% | 64% |
| 5 | 7 | 58% | 50% |

TABLE II. Precision and Recall without Phong Shading

| Category | No. of relevant image retrieved | Precision (%) | Recall (%) |
|---|---|---|---|
| 1 | 04 | 33.3% | 28.57% |
| 2 | 01 | 8.3% | 7.1% |
| 3 | 09 | 75% | 64.5% |
| 4 | 06 | 50% | 42% |
| 5 | 05 | 41% | 35% |

As per analysis of our result it is concluded that phong shaded precision and recall value is much more. Shaded image produce better color texture and edge density features value. Due to phong shaded concept more relevant images are retrieved. But in without phong shaded image more irrelevant images are found and less number of relevant images.

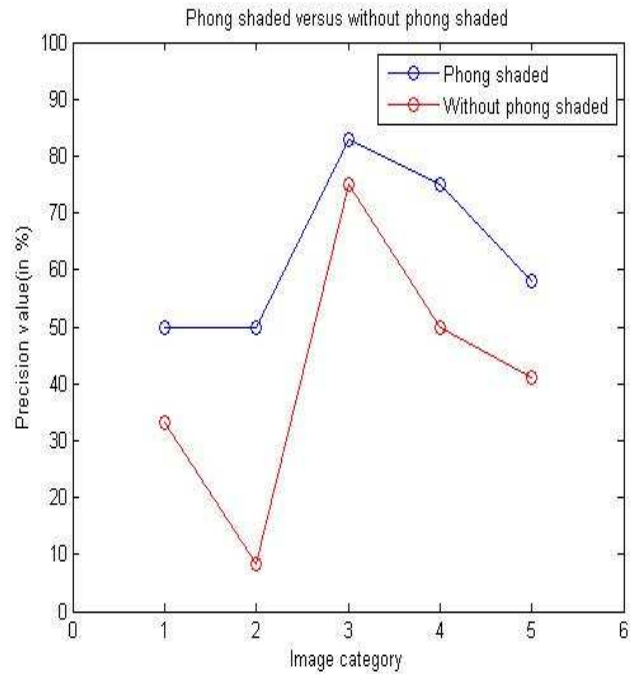

Figure8 :Precision for phong shaded versus without phong shade

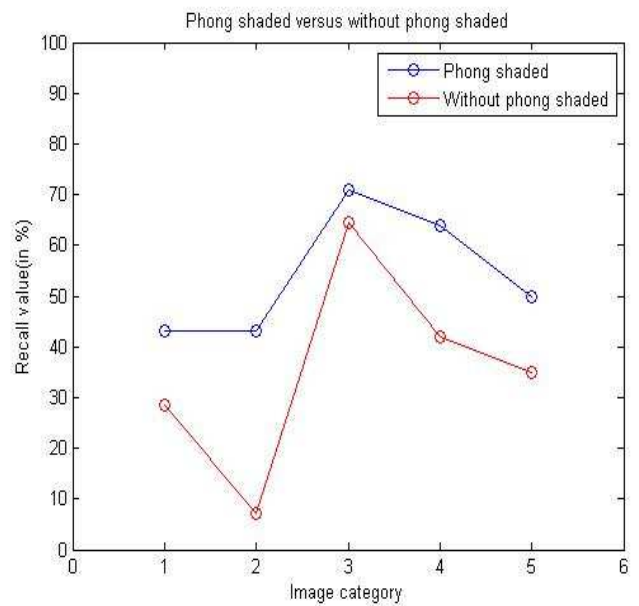

Figure9:Recall for phong shaded versus without phong shade

## VIII. CONCLUSION

Different features are used for retrieval of images such as Image color, texture and edges. The results demonstrate that these features are useful in retrieving similar images when a query image is presented. As per analysis of our result it is concluded that phong shaded image produce better color texture and edge density features value and we get more relevant images & less irrelevant images. But in without phong shaded image more irrelevant images are found & less number of relevant images. Experimental evaluations of the proposed framework have shown that it is effective and improves the retrieval performance of CBIR systems significantly.

We have designed our CBIR System using MATLAB (R2007b), to accomplish this research work. We have evaluated our proposed work on Coral dataset. The experimental result demonstrates the encouraging performance.

## IX. FUTURE WORK

The retrieval efficiency and timing performance can be further increased if the image collection is used best shading technique. With that, the image with high similarities in the feature space will be group together and result a smaller search space. This will greatly enhance the search time and precision and recall values. The results will give an indication of the possibilities presented by using a images based on color, texture properties in the context of an image retrieval system.

Results from a preliminary test of the systems indicate that the functionality provided in the framework described in this paper might provide useful. It is believed that the full experiment, which is in progress at the time of writing, will provide more conclusive evidence. Much work needs to be done in future.

**AUTHORS PROFILE**

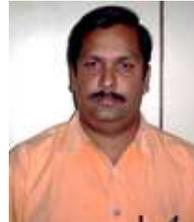

**Dr. Sanjeev Jain** is working as a Professor in Department of Computer Science and Engineering at LNCT Collage, Bhopal, India. He has completed M.Tech. From IIT Delhi and Ph.D. from Barkatullah University, He is having teaching experience of 21 Years with area of specialization in Image processing and Mobile Adhoc Networks. He has guided 5 Ph.D students and 41 M.Tech dissertation. He published about 70 papers in National/International Journal & Conference proceedings.

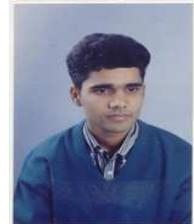

**Uday Pratap Singh** is working as a Assistant Professor in Department of Computer Science and Engineering at LNCT Collage, Bhopal, India. He has Completed his M.Sc.(Maths & Computing) from IIT Guwahati, and Persuing PhD from Barkatullah University. He published 10 papers in National/International Journal & Conference proceedings. His fields of interests are Computer Graphics and Image Processing.

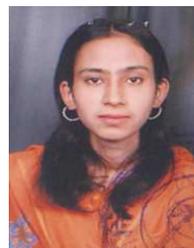

**Gulfishan Firdose Ahmed** was born in Bhopal, India. Now she is a M.Tech(CSE) Student in the Department of Computer Science & Engineering at the LNCT Bhopal in RGPV University. She published 2 papers in National/International Conference proceedings.. Her research fields include Computer Graphics and Image Processing.